\documentclass[12pt,final]{elsart}

  \usepackage{latexsym,bm,amsmath,amssymb,amsfonts}
  \usepackage{epsfig,graphics,graphicx}
  \usepackage{slashed}

  \long\def\comment#1{ }

  \newcommand{\beq}{\begin{eqnarray}}
  \newcommand{\eeq}{\end{eqnarray}}
  
 \def\simge{\mathrel{%
   \rlap{\raise 0.511ex \hbox{$>$}}{\lower 0.511ex \hbox{$\sim$}}}}
\def\simle{\mathrel{
   \rlap{\raise 0.511ex \hbox{$<$}}{\lower 0.511ex \hbox{$\sim$}}}}

\begin{document}

\begin{frontmatter}

\parbox[]{16.0cm}{ \begin{center}
\title{Towards thermalization in heavy--ion collisions: \\CGC meets the 2PI formalism}

\author{Yoshitaka Hatta and Akihiro Nishiyama }

\address{  Graduate School of Pure and Applied Sciences, University
of Tsukuba, \\Tsukuba, Ibaraki 305-8571, Japan
}


\end{center}

\begin{abstract}
We propose to apply the two--particle irreducible (2PI) formalism to the problem of thermalization in heavy--ion collisions in the Color Glass Condensate (CGC) picture. We consider the 2PI effective action to three loops and derive a set of coupled equations for the classical Yang--Mills field and the quantum fluctuations in the boost invariant coordinate system. The initial condition and the relation to previous works are also discussed.
\end{abstract}
}

\end{frontmatter}

\section{Introduction}

Ultra-relativistic heavy--ion collision experiments at BNL--RHIC and at the CERN--LHC offer a unique opportunity to give us a glimpse of QCD matter under extreme conditions and the possible formation of the quark--gluon plasma (QGP) \cite{Antinori:2011us}. This matter is intrinsically non-stationary, and understanding its expansion (`the little bang') bears a close parallel with tracing the history of the early universe at the dawn of the QCD epoch. Experimental observables related to collective flow and particle  correlations put strong constraints on the late stage of the evolution which appears to be well described by hydrodynamics \cite{Teaney:2009qa}. On the other hand, much less is understood about the early stage of the evolution, say, within 1 ${\rm fm}/c$ after the collision when the equation of state is not yet established, and one thus has to deal with the specific dynamics of the microscopic degrees of freedom in QCD.

The color glass condensate (CGC) \cite{Gelis:2010nm} is arguably the most solid framework to date to describe the very early stage of the nucleus collision when the coherent, strong color fields of the incoming nuclei are instantly liberated. What happens slightly after is vastly more complex. To first approximation, the fields of the CGC continue to evolve according to the classical Yang--Mills equation in the forward light--cone \cite{Kovner:1995ja}. While this picture is  useful for computing certain observables,  it alone has little to do with the problem of thermalization. The latter inevitably requires the consideration of quantum fluctuations which, fueled by the decay of the classical field, grow explosively and drastically change the fate of the evolution. In the context of CGC, the importance of the fluctuations was first realized in \cite{Romatschke:2005pm}, though they were introduced only heuristically, as random variables. A more proper treatment of the fluctuations as dynamical quantum fields has been implemented in \cite{Fukushima:2006ax,Dusling:2010rm,Dusling:2011rz}. The current state of the art \cite{Dusling:2010rm,Dusling:2011rz,Epelbaum:2011pc} is that one has a well--defined resummation scheme (originally devised for a scalar theory \cite{Son:1996zs}) in which the classical solution is dressed up by quantum fluctuations to all orders. This will be reviewed in the discussion section.

Meanwhile, over the past decades significant progress has been  made in the first--principle calculations of quantum field theories out of equilibrium,
which goes under the name of the two--particle irreducible (2PI) formalism, or more generally, the $n$--particle irreducible ($n$PI) formalism \cite{Berges:2004yj}. The 2PI formalism was initially developed for zero and finite temperature field theories, and is based on the Cornwall--Jackiw--Tomboulis (CJT) effective action  $\Gamma[\phi,G]$ \cite{Cornwall:1974vz} which  \emph{ab initio} treats (the Green's function of) the quantum fluctuations $G$ on equal footing with the classical field $\phi$.
 When applied to the real--time evolution of nonequilibrium systems, the 2PI formalism can be naturally viewed as the field--theoretic  generalization \cite{Ivanov:1998nv} of the classic, self--consistent (`$\Phi$--derivable') method in quantum statistical mechanics \cite{kada,Baym:1962sx}.
A variety of nonequilibrium processes have  been studied in this framework mostly in the context of  scalar field theories  \cite{Berges:2000ur,Aarts:2001qa,Aarts:2002dj,Juchem:2003bi,Berges:2004ce,
Arrizabalaga:2005tf,Tranberg:2008ae,Nishiyama:2008zw}.
 In particular, it has been demonstrated  that, starting from an arbitrary initial condition far from equilibrium, the 2PI dynamics drives the system towards the \emph{quantum} equilibrium characterized by the Bose--Einstein distribution.
  On the other hand, there are not many applications to  gauge theories due partly to subtleties  in formal theory regarding gauge invariance and renormalization \cite{Arrizabalaga:2002hn,Carrington:2003ut,Reinosa:2007vi,Reinosa:2009tc}. So far,   practical simulations of non--Abelian gauge theories far from equilibrium have been limited to the so--called classical statistical approximation \cite{Berges:2007re,Berges:2008mr} (see, also, \cite{Kunihiro:2010tg}) which has the merit of allowing one to include 2PI diagrams to all orders in the loop expansion at weak coupling, but systematically neglects at each order certain quantum contributions which are necessary to achieve the quantum equilibration. The very same approximation is actually involved in the CGC--based approach  \cite{Dusling:2010rm,Dusling:2011rz} as we shall see in the discussion section.

 In this paper, we derive a set of coupled equations describing the evolution of the classical color field \`a la CGC and its quantum fluctuations from the 2PI action to three loops in the boost invariant coordinate system.  By truncating the expansion to fixed order, we do not include higher loop 2PI diagrams as in the classical statistical method, but we do not employ the classical approximation, either.  Our work is basically a marriage of the CGC approach and the 2PI formalism. While the latter has been developed, in part at least, with the motivation of studying the problem of thermalization in heavy--ion collisions, there does not seem to be a previous application of the formalism to the realistic setup of the collisions. We attempt to fill this gap.  [See \cite{Nishiyama:2010mn} for an earlier work in the flat coordinates in the absence of the classical field.]

The paper is organized as follows. In Section 2, we describe the initial stage of heavy--ion collisions as seen from the CGC viewpoint. In Section 3, we give a brief review of the 2PI formalism. We then  derive in Section 4 the nonequilibrium evolution equations in the coordinate space from the 2PI effective action. The results are very complicated already at two loops, so in Section 5 we make a simplifying assumption and discuss the equations in the momentum space along with the initial condition.  We then conclude in Section 6 by comparing our results with the previous CGC approach \cite{Dusling:2010rm,Dusling:2011rz} from a diagrammatic point of view.

\section{Gluodynamics in the $A^\tau=0$ gauge}

  The matter created in the central region of ultra-relativistic heavy--ion collisions at RHIC and at the LHC is almost baryon--free and nearly boost invariant. To first approximation, one may describe such a system as  purely gluonic matter with strict boost invariance. A convenient choice of the coordinates is then the `$\tau$--$\eta$' coordinates defined by
 \beq
 \tau=\sqrt{t^2-(x^3)^2}\,, \qquad \eta = \tanh^{-1} \frac{x^3}{t}\,, \qquad x_\perp = (x^1,x^2)\,.
  \eeq
   The `proper time' $\tau$  plays the role of ordinary time $t$, and $\eta$ is the  rapidity.
  The components of the momentum in these coordinates are
\beq
p_\tau=\frac{1}{\tau}(tp^0 -x^3p^3)\,,\qquad p_\eta = x^3 p^0 -tp^3\,, \qquad p_\perp = (p_1,p_2)\,.
\eeq
  Boost invariance means that observables are independent of $\eta$, and the conjugate (dimensionless) momentum $p_\eta$ is conserved. An immediate consequence of this is that, around midrapidity where $x^3\approx 0$ and $t\approx \tau$, the longitudinal momentum decreases as $p_3 \approx p_\eta/\tau$.

We shall be interested in the $\tau$--evolution of the gluonic matter in the forward light--cone $\tau>0$. A popular choice of gauge is
\beq
A^\tau =A_\tau= \frac{1}{\tau}(x^-A^+ + x^+ A^-)=0\,,
\eeq
which is the analog of the temporal axial gauge $A^0=0$ in the ordinary  coordinates.
The Yang--Mills action in this gauge is
\beq
S_{YM}=\int \tau d\tau d\eta d^2x_\perp  \left[ \frac{1}{2\tau^2}(\partial_\tau A_\eta)^2
+\frac{1}{2}(\partial_\tau A_i)^2 -\frac{1}{2\tau^2}F_{\eta i}F_{\eta i} -\frac{1}{4} F_{ij}F_{ij} \right]\,,  \label{ya}
\eeq
where sums over the transverse directions $i,j=1,2$ and the color indices $a,b,..=1,..,N_c^2-1$ are understood. We have lowered the Lorentz indices on fields, which we shall do throughout this paper.
Introducing the notation  $x^\alpha=(\eta, x_\perp)$ and the spatial metric $\gamma_{\alpha\beta}\equiv {\rm diag}\, (\tau^2,1,1)$, the action can be compactly written as
\beq
S_{YM}=\int d\tau d\eta d^2x_\perp \sqrt{\gamma}  \left[ \frac{1}{2}\gamma^{\alpha\beta}\partial_\tau A_\alpha \partial_\tau A_\beta
-\frac{1}{4} \gamma^{\alpha\beta}\gamma^{\gamma\delta}F_{\alpha\gamma}F_{\beta\delta} \right]\,,  \label{ya}
\eeq
where $\gamma=\tau^2$ is the determinant of $\gamma_{\alpha\beta}$.

 The classical equation of motion following from the action (\ref{ya}) is
  \beq
\frac{1}{\sqrt{\gamma}}\frac{\delta S_{YM}}{\delta A_\alpha} = - \frac{1}{\sqrt{\gamma}}\partial_\tau \left(\sqrt{\gamma}\gamma^{\alpha\beta}\partial_\tau A_\beta \right)+ \gamma^{\alpha\beta}\gamma^{\gamma\delta}D_\gamma F_{\delta\beta}=0\,,  \label{ym}
 \eeq
  or in components,
 \beq
&& - \frac{1}{\tau} \partial_\tau \left(\frac{1}{\tau}\partial_\tau A_\eta\right) +\frac{1}{\tau^2}D_i F_{i\eta} =0\,, \nonumber \\
&&  -\frac{1}{\tau} \partial_\tau \left(\tau \partial_\tau A_i\right) + \left(\frac{1}{\tau^2} D_\eta F_{\eta i} +D_j F_{ji}\right) =0\,.
 \label{cov}
 \eeq

This should be supplemented with the Gauss's law constraint
\beq
D_\alpha E^\alpha= D_i E^i + D_\eta E^\eta=0\,,  \label{gauss}
\eeq
  where the conjugate momenta $E^\alpha=\sqrt{\gamma}\gamma^{\alpha\beta}\partial_\tau A_\beta$ are
  \beq
  E^i=\tau \partial_\tau A_i\,, \qquad E^\eta = \frac{1}{\tau}\partial_\tau A_\eta\,.
  \eeq
The energy density at the classical level is given by
\beq
  {\mathcal E} =  T_{\tau\tau} &=& \frac{1}{2}\gamma^{\alpha\beta}\partial_\tau A_\alpha \partial_\tau A_\beta
+\frac{1}{4} \gamma^{\alpha\beta}\gamma^{\gamma\delta}F_{\alpha\gamma}F_{\beta\delta} \nonumber \\ &=& \frac{1}{2\tau^2}(\partial_\tau A_\eta)^2 + \frac{1}{2}(\partial_\tau A_i)^2 + \frac{1}{2\tau^2}F_{\eta i}F_{\eta i} +\frac{1}{4}F_{ij}^2\,.
\eeq
However $E\equiv \int \tau d\eta d^2x_\perp {\mathcal E}$ is not constant in $\tau$ because the metric $\gamma_{\alpha\beta}$ depends on $\tau$. The continuity equation reads
\beq
\partial_\tau \left(\int \tau d\eta d^2x_\perp {\mathcal E}\right) =-\tau^2 \int d\eta d^2x_\perp T^{\eta\eta}\,,
\eeq
where the `pressure' in the $\eta$--direction is
\beq
\tau^2T^{\eta\eta}= -\frac{1}{2\tau^2}(\partial_\tau A_\eta)^2 + \frac{1}{2}(\partial_\tau A_i)^2 + \frac{1}{2\tau^2}F_{\eta i}F_{\eta i} -\frac{1}{4}F_{ij}^2\,.
\eeq

  In the framework of the color glass condensate (CGC),  the solution $A_\alpha$ of (\ref{ym}) represents the strong color field liberated from the colliding nuclei. At early stages, it is parametrically of order $A_\alpha \sim Q_s/g$ where $Q_s$ is the so--called saturation momentum. Immediately after the collision at $\tau=0$, it can be written as
   \cite{Kovner:1995ja}
 \beq
 &&A_{i}={\mathcal A}^1_{i} +{\mathcal A}^2_i\,, \qquad A_\eta=0\,, \\
 && \tau \partial_\tau A_i =0\,, \qquad \frac{1}{\tau}\partial_\tau A^a
 _\eta= -gf_{abc}{\mathcal A}_i^{1b} {\mathcal A}_i^{2c}\,, \label{ini}
 \eeq
 where  ${\mathcal A}^{1}$ and ${\mathcal A}^{2}$ are the color fields of the projectile (nucleus 1) and the target (nucleus 2), respectively, before the collision. They are `pure gauge' in the transverse plane $\partial_i {\mathcal A}_j -\partial_j {\mathcal A}_i -ig[{\mathcal A}_i, {\mathcal A}_j]=0$ and are related to the color charge density of valence partons $\rho$ via  $\partial_i {\mathcal A}_i^{1,2} = \rho^{1,2}$.
 In the McLerran--Venugopalan model \cite{McLerran:1993ni}, one calculates observables ${\mathcal O}[A[{\mathcal A}]]$ (such as the gluon multiplicity and energy density) using the solution $A$ to the classical equation of motion which in turn is a functional of the initial fields ${\mathcal A}^{1,2}$. One then averages  over ${\mathcal A}^{1,2}$, or equivalently, $\rho^{1,2}$
\beq
\langle {\mathcal O}\rangle = \int {\mathcal D}\rho^1{\mathcal D}\rho^2\, W[\rho^1]W[\rho^2]{\mathcal O}[{\mathcal A}[\rho]]\,, \label{asin}
\eeq
where the Gaussian weight functional
\beq
W[\rho]\sim \exp\left(-\frac{g^2}{2\mu^2}\int d^2x_\perp \rho_a(x_\perp)\rho_a(x_\perp)\right)\,,
\eeq
with $\mu \sim {\mathcal O}(Q_s)$ takes into account the randomness of the color charge distribution in the transverse plane of a heavy nucleus.\\

  Expanding the action around the classical solution $A_\alpha \to A_\alpha + a_\alpha$, one finds the linearized equation of motion for the quantum fluctuations $a_\alpha$
 \beq
 &&  -\frac{1}{\sqrt{\gamma}} \partial_\tau \left( \sqrt{\gamma}\gamma^{\alpha\beta} \partial_\tau a_\beta \right)
 +\left(D^2 \gamma^{\alpha\beta} -\gamma^{\alpha \gamma}\gamma^{\beta \delta}D_\gamma D_\delta -2ig \gamma^{\alpha\gamma}\gamma^{\beta \delta}F_{\gamma\delta} \right) a_\beta
 \nonumber \\
 &&  =  -\frac{1}{\sqrt{\gamma}} \partial_\tau ( \sqrt{\gamma}\gamma^{\alpha\beta} \partial_\tau a_\beta )
 +\left( D^2 \gamma^{\alpha\beta} +\gamma^{\alpha\gamma}\gamma^{\beta\delta}(D_\gamma D_\delta -2D_\delta D_\gamma) \right)a_\beta \nonumber \\
  &&  =0\,,   \label{fluct}
 \eeq
 where $D^2 = \gamma^{\alpha\beta}D_\alpha D_\beta$ and  $D_\alpha = \partial_\alpha -igA_\alpha$ is the covariant derivative constructed from the classical solution. [Note that $[D_\alpha,D_\beta]=-igF_{\alpha\beta}$.]
 The interaction terms in the action read
 \beq
 S_{int}=\int d\tau d\eta d^2x_\perp \sqrt{\gamma}\left({\mathcal L}_3 + {\mathcal L}_4 \right)\,,
 \eeq
 with the three-- and four--point vertices
 \beq
{\mathcal L}_3+{\mathcal L}_4=-gf_{abc}\gamma^{\alpha\gamma}\gamma^{\beta\delta}(D_\alpha a_\beta)^a a_\gamma^b a_\delta^c
  -\frac{g^2}{4}f_{abc}f_{ab'c'}\gamma^{\alpha\gamma}\gamma^{\beta\delta}a_\alpha^b a_\beta^c a_\gamma^{b'}a_\delta^{c'}\,.  \label{ver}
 \eeq
 For later use, we introduce a shorthand notation
   \beq
 C_{bb',cc'} \equiv f_{abc}f_{ab'c'}+f_{abc'}f_{ab'c}\,,
 \eeq
 which is symmetric under $b\leftrightarrow b'$, $c \leftrightarrow c'$ and $(bb') \leftrightarrow (cc')$. With this definition,
  (\ref{ver}) can be written as
\beq
{\mathcal L}_3+{\mathcal L}_4 &=&  -gf_{abc}\gamma^{\alpha\gamma}\gamma^{\beta\delta}(\partial_\alpha a_\beta)^a a_\gamma^b a_\delta^c   -\frac{g^2}{2}C_{bb',cc'} \gamma^{\alpha\gamma}\gamma^{\beta\delta} A_\alpha^{b'} a_\beta^{c'} a_\gamma^b a_\delta^{c} \nonumber \\ && \qquad \qquad -\frac{g^2}{8}C_{bb',cc'} \gamma^{\alpha\gamma}\gamma^{\beta\delta}
  a_\alpha^{b'} a_\beta^{c'} a_\gamma^{b}a_\delta^{c}  \,. \label{ver2}
\eeq


\section{The two--particle irreducible (2PI) formalism}

The equations for the classical field (\ref{ym}) and the fluctuations (\ref{fluct}) by themselves do not contain the essential dynamics that drives the system to quantum equilibration. In order to go beyond, one has to consider the nonlinear interaction of the fluctuations shown in (\ref{ver}) and its backreaction to the classical field. As explained in the Introduction, a powerful framework to discuss these issues is the two--particle irreducible (2PI) formalism which we shall review in this section.

Real--time phenomena are best formulated in the Keldysh formalism which involves the doubling of the field degrees of freedom  living on the closed time path (CTP)---a union of the forward $[\tau_0,\infty]$ and backward $[\infty, \tau_0]$ time branches. [$\tau_0 \approx 0$ is the initial time.]  We define the propagator of the fluctuation
 \beq
 G^{ab}_{\alpha\beta}(x,y) = \langle \mbox{T}_C\{a_\alpha^a (x) a_\beta^b (y) \} \rangle\,,   \label{pro}
   \eeq
   where the symbol T$_C$ denotes path--ordering along the CTP and the brackets $\langle ... \rangle$ denote averaging over the density matrix initially prepared at $\tau=\tau_0$.  The effective action of the classical field $A$ and the quantum fluctuation $G$ is given by
\beq
\Gamma[A,G]=S_{YM}[A] +\frac{i}{2}{\rm tr} \ln G^{-1} + \frac{i}{2}{\rm tr}\, G^{-1}_0[A] G +\Gamma_2[A,G]\,,  \label{act}
\eeq
 where $\Gamma_2$ is the sum of 2PI diagrams constructed from the exact propagators (\ref{pro}) and the bare interaction vertices (\ref{ver}).  $G_0$ is the  propagator in the presence of the classical background $A$ and satisfies (c.f. (\ref{fluct}))
  \beq
 && i\left(\frac{1}{\sqrt{\gamma}} \partial_\tau \left( \sqrt{\gamma}\gamma^{\alpha\beta} \partial_\tau  \right)
 -\bigl(D^2 \gamma^{\alpha\beta} +\gamma^{\alpha\gamma}\gamma^{\beta\delta}(D_\gamma D_\delta -2D_\delta D_\gamma)  \bigr)\right)^{ab} G^{bc}_{0,\beta\epsilon}(x,x') \nonumber \\
 && \qquad \qquad \qquad \qquad  = \delta^\alpha_{\epsilon} \delta^{ac} \frac{1}{\sqrt{\gamma}}\delta_C(\tau-\tau')\delta(\eta-\eta')\delta^2(x_\perp-x'_\perp) \,,   \label{quad}
 \eeq
 where the contour--delta function is defined by
 \beq
 \delta_{C}(\tau-\tau') = \begin{cases} \delta (\tau-\tau')\,, & \tau,\tau'\in [\tau_0,\infty]\,,\\
 -\delta(\tau-\tau')\,, & \tau,\tau'\in [\infty,\tau_0]\,. \\
 \end{cases}
 \eeq

 The equations of motion for $A$ and $G$ are derived from the stationary conditions
 \beq
 \frac{\delta \Gamma}{\delta A}=0\,, \label{eq1}
 \eeq
 and
 \beq
 \frac{\delta \Gamma}{\delta G}=0\,. \label{eq2}
 \eeq
 To lowest order, (\ref{eq1}) coincides with the Yang--Mills equation (\ref{ym}). But it receives quantum corrections (backreaction) from the fluctuation $G$ as we shall soon see. Likewise, $G$ deviates from $G_0$ due to the self--interaction of the fluctuations.
 (\ref{eq2}) may be written in a more familiar form (suppressing indices)
    \beq
G=G_0 +G_0\Pi G \,,  \label{green}
\eeq
 where $\Pi (x,x')\equiv 2i\delta \Gamma_2/\delta G(x,x')$ is the self--energy which has both the  local (tadpole) part proportional to $\delta^4(x-x')$ and the nonlocal part. Since the local part receives contributions only at one--loop and can be easily separated out, in the following we shall use the notation $\Pi(x,x')$ only for the nonlocal part.

One of the merits of the 2PI formalism is the absence of the so--called secular terms \cite{Berges:2004yj}. The secular terms appear ubiquitously in perturbative approaches to evolution equations in time, and have to be resummed by some means. Since the equations  (\ref{eq1}) and (\ref{eq2}) in the 2PI formalism are those for the \emph{exact} Green's function which resums infinitely many 1PI diagrams, the secular terms do not arise at any stage. The price to pay is that one has to solve nonlinear equations. This is however straightforward to do
 numerically as an initial value problem, thanks to the manifest causal structure of the equations.

\begin{figure}
  \begin{center}
      \includegraphics[width=130mm]{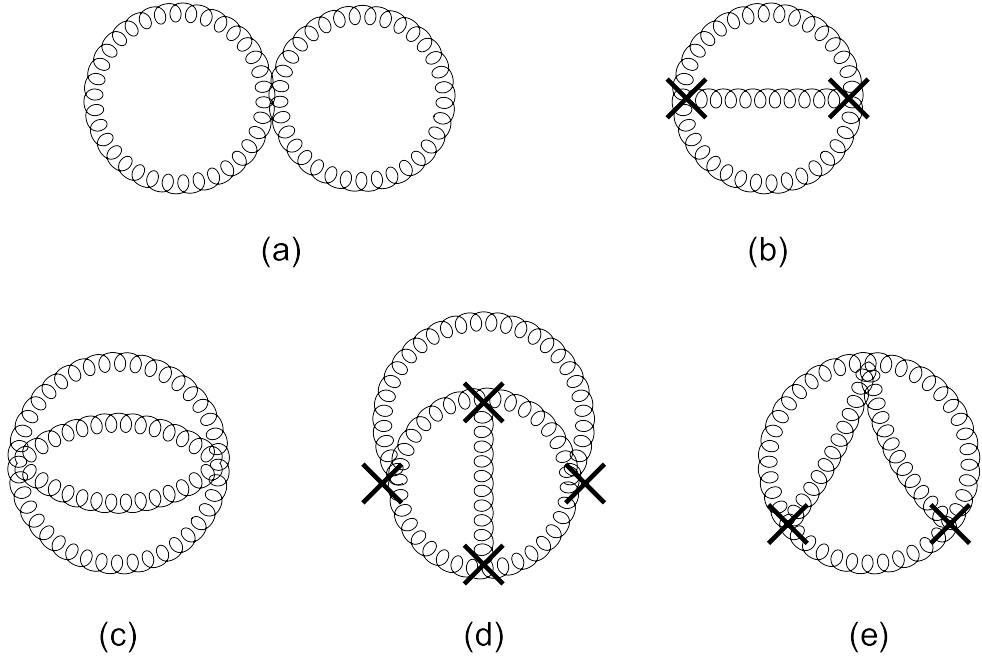}
  \end{center}
  \caption{2PI diagrams up to three loops in the presence of the classical field. Crosses indicate the insertion of the covariant derivative.}
  \label{f1}
\end{figure}

 In this paper, we assume weak coupling $g\ll 1$ and consider $\Gamma_2$ up to three loops.
 The relevant diagrams are depicted in Fig.~1. In these diagrams, crosses represent the insertion of the covariant derivative $D[A]=\partial -ig A$ where $A$ is the classical field.   At early times when $A\sim {\mathcal O}(1/g)$, the two terms in the covariant derivative are of the same order in the coupling.  In this regime, the evolution is driven by Diagram \ref{f1}b which describes the decay (decoherence) of the classical field, the scattering of quantum fluctuations off the classical field and the $1\leftrightarrow 2$ processes of the fluctuations. [The last process was previously considered in \cite{Nishiyama:2010mn} in the flat metric.] With increasing $\tau$, the magnitude of $A$ decreases, and so do the diagrams with insertions of $A$'s. When $A\sim {\mathcal O}(g^0)$, they become of the same order as the three--loop diagrams.

Comments are in order concerning gauge invariance. We first assure the reader that the residual gauge symmetry in the $A^\tau=0$ gauge will be respected in the evolution equations to be derived below, in the sense that they transform covariantly under residual ($\tau$--independent) gauge transformations.
 However, there are potential problems regarding more general ($\tau$--dependent) gauge transformations:
   By truncating the loop expansion at a fixed order in the coupling, one cannot guarantee the full gauge invariance of the 2PI effective action.\footnote{We note, however, that the problem at hand (gluodynamics in the $\tau-\eta$ coordinates) has never been discussed outside the $A^\tau=0$ gauge to our knowledge.}  This is a well--known problem of the 2PI approach and the general solution is not known. Still one can rely on the idea of ``controlled gauge invariance" \cite{Arrizabalaga:2002hn,Carrington:2003ut} which states that the gauge dependence appears at  higher order than the truncation order for the propagator, and at twice the order of truncation for physical quantities (pressure, etc.). Moreover, in QED  there exist generalized Ward identities for the correlation functions which save renormalizability \cite{Reinosa:2007vi,Reinosa:2009tc}. It would be very important to study these issues in the present context. We leave such a study for future work.

Before leaving this section we  make a technical observation which will greatly simplify the subsequent analysis. The equations shown in the previous section are cluttered with factors of the metric tensor $\gamma_{\alpha\beta}$. One can eliminate them by introducing a  dimensionful coordinate $\zeta \equiv  \tau \eta$, and accordingly, the following rescaling
\beq
A_\eta = \tau A_\zeta\,, \qquad a_\eta = \tau a_\zeta\,, \qquad  \partial_\eta = \tau \partial_\zeta\,. \label{rescale}
\eeq
 The equation of motion (\ref{cov}) then becomes
\beq
&& \frac{1}{\sqrt{\gamma}} \frac{\delta S_{YM}}{(\delta A_\eta)/\tau} = - \left(\partial_\tau^2 A_\zeta + \frac{1}{\tau}\partial_\tau A_\zeta -\frac{A_\zeta}{\tau^2} -D_i F_{i\zeta}\right) \,, \nonumber \\
&& \frac{1}{\sqrt{\gamma}} \frac{\delta S_{YM}}{\delta A_i} = - \left(\partial_\tau^2 A_i + \frac{1}{\tau} \partial_\tau A_i -D_\zeta F_{\zeta i} -D_j F_{ji}\right)\,.  \label{hen}
\eeq
Note that one should not do the rescaling (\ref{rescale}) {\it before} functionally differentiating the kinetic term of the action because it interferes with the $\tau$--derivative.
 Let us label the new set of spatial coordinates  $x^I=(\zeta,x_\perp)$, $\partial_I=(\partial_\zeta, \partial_\perp)$ by capital letters $I,J,..$. By slight abuse of notation,  we denote the two equations in (\ref{hen}) as
 \beq
\frac{\delta S_{YM}}{\delta A_I} =   -\left(\partial_\tau^2 A_I + \frac{1}{\tau}\partial_\tau A_I -\delta_{I\zeta}\frac{A_I}{\tau^2} -D_J F_{JI}\right)\,, \label{last}
  \eeq
   keeping the above caveat in mind. Eq.~(\ref{last}) suggests that in the coordinate system $(\zeta, x_\perp)$ the equations (\ref{eq1}) and (\ref{eq2}) look as if we were working in the Cartesian coordinate system. Indeed, the interaction vertices (\ref{ver2}) take the same form as in flat space
 \beq
S_{int} = \int_C d\tau d^3x \Bigl( -gf_{abc}(\partial_I a_J)^a a_I^b a_J^c   -\frac{g^2}{2}C_{bb',cc'}  A_I^{b'} a_J^{c'} a_I^b a_J^{c}  -\frac{g^2}{8}C_{bb',cc'}
  a_I^{b'} a_J^{c'} a_I^{b}a_J^{c} \Bigr)  \,, \label{ver2}
\eeq
 where $d^3x = d\zeta d^2x_\perp$ and the subscript $C$ means that the $\tau$--integral is performed along the CTP.
 Similarly, the equation for the Green's function (\ref{fluct}) becomes
    \beq
&& i\left[\left(\partial_\tau^2+\frac{1}{\tau}\partial_\tau -\frac{\delta_{I\zeta}}{\tau^2}\right) \delta_{IJ} -(D^2 \delta_{IJ}+D_ID_J-2D_JD_I)\right]^{ab}G^{bc}_{0,JK}(x,x')\nonumber \\
 &&  \qquad \qquad \qquad \qquad =\delta_{IK} \delta^{ac} \delta_C(\tau-\tau')\delta^3(x_I-x'_I) \,, \label{fluct2}
\eeq
 where $D^2=D_ID_I$ and we have also rescaled the Green's functions $G_{\eta\eta}=\tau \tau' G_{\zeta\zeta}$, $G_{\eta i}=\tau G_{\zeta i}$, $G_{i\eta}=\tau'G_{i\zeta}$.
Since (\ref{ver2}) does not involve $\tau$--derivatives, no subtlety arises when taking the functional derivative of diagrams constructed from these vertices. Thus the use of the coordinates $x^I$ leads to  the simplest derivation and representation of the evolution equation without any complications from the metric tensor. The only cautionary remark is that one has to remember which value of $\tau$ is used in the  rescaling $\zeta = \tau \eta$ in each of the covariant derivatives $D_\zeta$ that appear in the equations.

\section{Nonequilibrium evolution equations}

In this section we explicitly write down the evolution equations (\ref{eq1}) and (\ref{eq2}) using the two--loop effective action. The contributions from the three--loop action are complicated and thus relegated to the Appendix.
  Actually, at the level of the derivation of the evolution equations we do not have to  assume boost invariance, and the results in this section are valid for generic backgrounds $A_I$ depending on all coordinates $(\tau,\eta, x_\perp)$. Assumptions on the $(\eta,x_\perp)$--dependence come later, when we specify the initial conditions and make approximations to solve these equations.

\subsection{Evolution equations in the coordinate space}

 Let us now evaluate the functional derivative in (\ref{eq1}).  Consider first the term
\beq
\frac{\delta}{\delta A_I} \left(\frac{i}{2}{\rm tr}\, G_0^{-1}G\right) \sim \frac{1}{2}\frac{\delta}{\delta A_I} \Bigl({\rm tr}\,  (D^2 \delta_{IJ}+D_ID_J-2D_JD_I)^{ab}G^{ba}_{JI}\Bigr)\,,
\eeq
 where the symbol $\sim$ means we only keep terms which contain $A$.
Explicitly,
\beq
\frac{1}{2}{\rm tr}\, (D^2)^{ab}G_{JJ}^{ba} \sim \frac{gf_{abc}}{2}\int_C dx \left(\partial_L (A_L^b G_{JJ}^{ca}) + A_L^b\partial_L G_{JJ}^{ca} + gf_{cde}A_L^b
A_L^d G_{JJ}^{ea}\right)_{y=x}\,,
\eeq
\beq
\frac{1}{2}{\rm tr}\, (D_ID_J)^{ab}G_{JI}^{ba} \sim \frac{gf_{abc}}{2} \int_C dx \left( \partial_I(A_J^b G_{JI}^{ca})+A_I^b \partial_J G_{JI}^{ca} +g f_{cde}A_I^bA_J^d G_{JI}^{ea} \right)_{y=x}\,. \label{rhs}
\eeq

Differentiating, we obtain
\beq
\frac{\delta}{\delta A_I^a} \left(\frac{1}{2}{\rm tr}\, (D^2)^{ab}G_{JJ}^{ba} \right) &=&  \frac{g}{2}f_{bac}\bigl(\partial^x_I G_{JJ}^{cb}(x,y) -\partial^x_IG_{JJ}^{cb}(y,x)\bigr)_{y=x} \nonumber \\
&& \qquad \qquad +\frac{g^2}{2}(f_{bac}f_{cde}+f_{bdc}f_{cae})A_I^d G_{JJ}^{eb}(x,x) \nonumber \\
&=&
\frac{g}{2}f_{abc} D_{xI}^{be}
 (G_{JJ}^{ec}(x,y) +  G_{JJ}^{ce}(y,x))_{y=x}\,,
 \eeq
 where $D_{xI}^{be}=\partial^x_I\delta^{be}+gf_{bde}A_I^d(x)$ and similarly,
\beq
\frac{\delta}{\delta A_I^a} \left(\frac{1}{2}{\rm tr}\, (D_ID_J)^{ab}G_{JI}^{ba} \right) &=&
\frac{g}{2}f_{bac}\bigl(\partial^x_J G_{JI}^{cb}(x,y)-\partial^x_J G_{IJ}^{cb}(y,x)\bigr)_{y=x}  \nonumber \\
 && \qquad \qquad +\frac{g^2}{2}(f_{bac}f_{cde}A_J^d G_{JI}^{eb} +f_{bdc}f_{cae}A_J^d G_{IJ}^{eb}) \nonumber \\
&=& \frac{g}{2}f_{abc} D_{xJ}^{be}
(G_{JI}^{ec}(x,y)
+G_{IJ}^{ce}(y,x) )_{y=x}\,.
\eeq
Noting that $G_{JI}^{ec}(x,y) =  G_{IJ}^{ce}(y,x)$, we find
\beq
\frac{\delta}{\delta A_I^a} \left(\frac{i}{2}{\rm tr}\, G_0^{-1}G\right)
= gf_{abc} \Bigl[ D_{xI}^{be}
 G_{JJ}^{ec}(x,y)  + D_{xJ}^{be}
 \bigl(G_{JI}^{ec}(x,y)- 2G_{IJ}^{ec}(x,y)\bigr) \Bigr]_{y=x} \,. \label{1}
\eeq

Next we calculate $\delta \Gamma_2/\delta A$.
To two loops, the relevant diagram is  Fig.~\ref{f1}b with the interaction vertices shown in (\ref{ver2})
\beq
\frac{\delta \Gamma_2}{\delta A_I^a} &=&-i \frac{(-ig^2)}{2} C_{ab,cd}a_I^{b}a_J^c a_J^{d} \int_C (-ig f_{lmn})(D_M a_N)^{l} a_M^m a_N^n\Big\arrowvert_{{\rm contractions}} \nonumber \\
&=& -\frac{g^2}{2}C_{ab,cd}\int_C dy \,V^y_{lmn,LMN}G_{IL}^{bl}(x,y)G_{JM}^{cm}(x,y)G_{JN}^{dn}(x,y)\,, 
\label{AA}
\eeq
 where we defined the (dressed) three--gluon vertex
 \beq
 V_{lmn,LMN}&=& -ig\Bigl\{ \delta_{LM}(2f_{lm'n}D^{m'm}_N+f_{lmn'}D^{n'n}_N  )  \nonumber \\
&&  \qquad + \delta_{LN}(-f_{lm'n}D^{m'm}_M-2f_{lmn'}D_M^{n'n} ) \nonumber \\
&& \qquad \qquad + \delta_{MN}(-f_{lm'n}D^{m'm}_L + f_{lmn'}D_L^{n'n} ) \Bigr\}\,.
\eeq
  In (\ref{AA}) and in the following, it is understood that each covariance derivative in $V$ acts only  on  one of the $G$'s  which has the same color index.
Combining Eqs.~(\ref{last}), (\ref{1}) and (\ref{AA}), one finds the equation of motion for the background in the presence of quantum fluctuations.
\beq
&&  \left(\partial_\tau^2 A_I + \frac{1}{\tau}\partial_\tau A_I -\delta_{I\zeta}\frac{A_I}{\tau^2} -D_J F_{JI}\right)^a \nonumber \\
 && \quad=gf_{abc} \Bigl[ D_{xI}^{be}
 G_{JJ}^{ec}(x,y) + D_{xJ}^{be}
 \bigl(G_{JI}^{ec}(x,y)- 2G_{IJ}^{ec}(x,y)\bigr) \Bigr]_{y=x} \nonumber \\
&& \qquad -\frac{g^2}{2}C_{ab,cd}\int_C dy \,V^y_{lmn,LMN}G_{IL}^{bl}(x,y)G_{JM}^{cm}(x,y)G_{JN}^{dn}(x,y).
\label{origin}
\eeq
At the three--loop order, there are additional contributions, (\ref{1d}) and (\ref{1e}), to the right hand side of this equation.

 \begin{figure}
  \begin{center}
      \includegraphics[width=160mm]{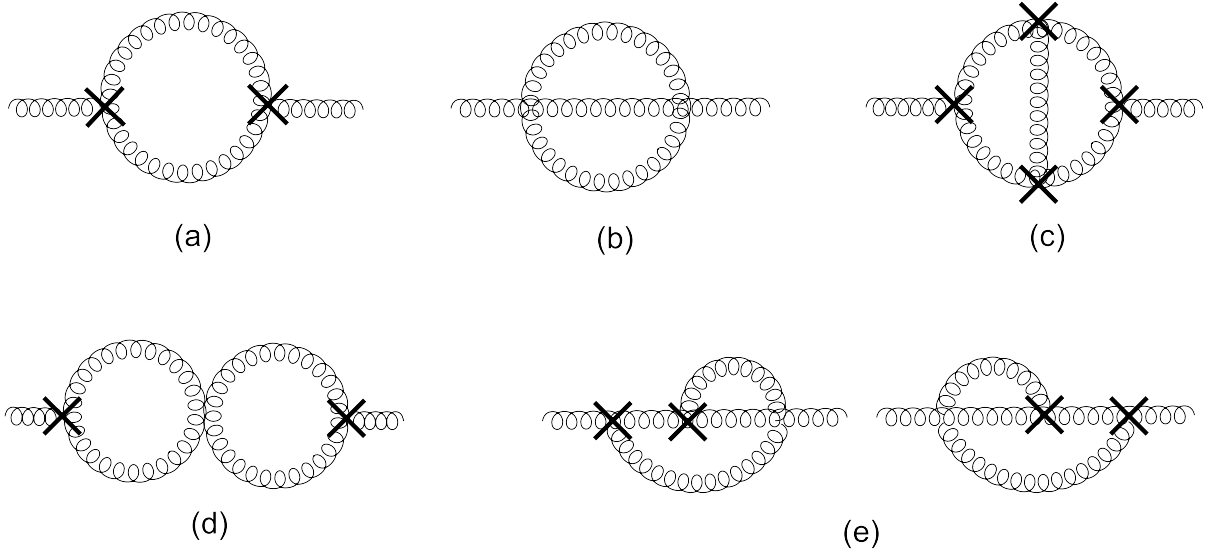}
  \end{center}
  \caption{The self energy diagrams up to two loops (from the three--loop effective action).}
  \label{f2}
\end{figure}

The equation for the Green's function (\ref{green}) reads
\beq
&& \left[\left( \partial_\tau^2  + \frac{1}{\tau}\partial_\tau  -\frac{1}{\tau^2} \delta_{I\zeta}\right)\delta_{IJ} -(D^2 \delta_{IJ}+D_ID_J-2D_JD_I)\right]^{ab}G^{bc}_{JK}(x,y) \nonumber \\
 &&  \qquad \qquad + g^2  \left( C_{ad,be} G^{de}_{IJ}(x,x) + \frac{1}{2} C_{ab,de}G^{de}_{MM} (x,x)\delta_{IJ} \right) G^{bc}_{JK}(x,y) \nonumber \\
 && \qquad  \qquad \qquad + i \int_C dz\,  \Pi_{IJ}^{ab}(x,z) G^{bc}_{JK}(z,y)
\nonumber \\
&& \qquad  =-i\delta^{ac}\delta_{IK} \delta_C(x-y)\,,    \label{new}
\eeq
where we have separated out the local part of the self energy coming from Fig.~\ref{f1}a.
   To one--loop (two--loop in $\Gamma_2$), the nonlocal part  comes from a single diagram  Fig.~\ref{f2}a and is given by
 \beq
 \Pi_{IJ}^{ab} (x,y) = \frac{1}{2}\, \overrightarrow{V}^x_{alm,ILM}G_{LL'}^{ll'}(x,y)G_{MM'}^{mm'}(x,y)\overleftarrow{V}^y_{bl'm',JL'M'}\,. \label{nonl}
 \eeq
To two loops ($\Gamma_2$ to three loops), there are five more self energy diagrams Fig.~\ref{f2}b--2e  which should be added to (\ref{nonl}). These are evaluated in (\ref{A})--(\ref{D}).\\

 It is important to notice that the above equations (\ref{origin}) and (\ref{new}) transform  covariantly under the residual ($\tau$--independent) gauge transformation  of the background
\beq
A_I \to UA_IU^\dagger +\frac{i}{g}U\partial_I U^\dagger \,,  \label{tran1}
\eeq
and the concomitant transformation of the fluctuations
\beq
&& a_I^a \to (Ua_IU^\dagger)^a = U^{ab}a_I^b\,, \label{tran2} \\
&& G_{IJ}^{ab}(x,y) \to U_x^{aa'}G_{IJ}^{a'b'}(x,y)(U_y^\dagger)^{b'b} = U_x^{aa'} U_y^{bb'} G_{IJ}^{a'b'}(x,y)\,, \nonumber
\eeq
so that the equations take the same form after a $U$--transformation.\footnote{The term $-A_\zeta/\tau^2$ on the left hand side of (\ref{origin}) may seem problematic at first. However, in the transformation
\beq
A_\zeta \to UA_\zeta U^\dagger +\frac{i}{g} U\partial_\zeta U^\dagger =
UA_\zeta U^\dagger +\frac{i}{g} \frac{1}{\tau}U\partial_\eta U^\dagger\,, \nonumber
\eeq
 the second term depends on $\tau$, and the inhomogeneous terms cancel among the first three terms of (\ref{origin}) so that
\beq
\partial_\tau^2 A_\zeta + \frac{1}{\tau}\partial_\tau A_\zeta -\frac{A_\zeta}{\tau^2}
\to U\left(\partial_\tau^2 A_\zeta + \frac{1}{\tau}\partial_\tau A_\zeta -\frac{A_\zeta}{\tau^2}\right) U^\dagger\,, \nonumber
\eeq
as it should. }

This can be checked by using the following identifies
\beq
 && f_{abc}U^{bb'}U^{cc'} = U^{aa'}f_{a'b'c'}\,, \qquad f_{abc}U^{aa'}U^{bb'}U^{cc'} =f_{a'b'c'}\,, \nonumber \\
 && C_{ab,cd}U^{bb'}U^{cc'}U^{dd'} = U^{aa'}C_{a'b',c'd'}\,,
\eeq
 which hold for a SU($N_c$) matrix $U^{ab}=(U^\dagger)^{ba}=(U^{-1})^{ba}$ in the adjoint representation. In fact, the two terms of the interaction Lagrangian (\ref{ver}) are already separately invariant under the transformations (\ref{tran1}) and (\ref{tran2}) which are a decomposition of  $A+a \to U(A+a)U^\dagger+\frac{i}{g}U\partial U^\dagger$. Hence the derivatives of  diagrams constructed from these vertices are expected to behave properly under the (residual) gauge transformation.

\subsection{The statistical and spectral functions}

 In practice, the equations (\ref{origin}) and (\ref{new}) are awkward  to handle because of the complicated time integral defined along the CTP. A standard trick to disentangle this is to rewrite the equations
 in terms of the so--called statistical ${\mathcal F}$ and spectral  $\rho$ parts of the Green's function\footnote{In the literature the statistical function is often denoted as $F$. Here we use the calligraphic letter in order to  avoid confusion with the field strength tensor $F_{IJ}$. For the spectral function we use the same letter $\rho$ as the source color charge (c.f., (\ref{asin})), but the distinction should be obvious from the context. }
 \beq
G(x,x')= \langle \mbox{T}_C\{a(x)a(x')\}\rangle &=& \frac{1}{2} \langle \{a, a'\} \rangle + \frac{1}{2}
 (\theta_C(\tau-\tau')-\theta_C(\tau'-\tau)) \langle [a,a'] \rangle \nonumber \\
 &\equiv & {\mathcal F} -\frac{i}{2} (\theta_C(\tau-\tau')-\theta_C(\tau'-\tau))\, \rho\,.  \label{def}
 \eeq
 Roughly,  the spectral function tells which states exist, while the statistical function tells how many particles are there in each state. The spectral function is related to the retarded Green's function $G_R$ as
 \beq
 G_R(x,x')=-i\theta(\tau-\tau')\rho(x,x')\,.
 \eeq
In the same way, we define the statistical and spectral parts of the nonlocal  self--energy
\beq
\Pi =  \Pi_{{\mathcal F}} -\frac{i}{2} (\theta_C(\tau-\tau')-\theta_C(\tau'-\tau))\, \Pi_\rho\,.
 \eeq
 From the definition (\ref{def}) in terms of the (anti-)commutator of fields, it follows that ${\mathcal F}(x,x')$ and $\rho(x,x')$ are real. That $\Pi_{\mathcal F}(x,x')$ and $\Pi_\rho(x,x')$ are also real is not immediately obvious, but can be understood from the identities
 \beq
 \Pi_{\mathcal F}&=& G_R^{-1}{\mathcal F} G_A^{-1}\,, \nonumber \\
 \Pi_\rho &=& i(G_R^{-1}-G_A^{-1})\,,
 \eeq
 where $G_A(x,x')=i\theta(\tau'-\tau)\rho(x,x')$ is the advanced propagator, and the fact  that $G_R$ and $G_A$ are purely imaginary in the coordinate space.

It is then a simple exercise to show that (\ref{origin}) can be rewritten as
\beq
 &&  \left(\partial_\tau^2 A_I + \frac{1}{\tau}\partial_\tau A_I -\delta_{I\zeta}\frac{A_I}{\tau^2} -D_J F_{JI}\right)^a \nonumber \\
 &&=
gf_{abc} \biggl( D_{xI}^{be}
 {\mathcal F}_{JJ}^{ec}(x,y) +D_{xJ}^{be} \bigl({\mathcal F}_{JI}^{ec}(x,y)
 -2{\mathcal F}_{IJ}^{ec}(x,y)\bigr) \biggr)_{y=x} \nonumber \\
  && \quad  +\frac{ig^2}{2} C_{ab,cd} \int^\tau_{\tau_0} d^4y \, V^y_{lmn,LMN} \biggl[ \rho_{IL}^{bl}(x,y){\mathcal F}_{JM}^{cm}(x,y){\mathcal F}_{JN}^{dn}(x,y) \nonumber \\
    && \qquad \qquad \qquad \qquad \qquad \qquad \qquad \qquad +({\mathcal F}\rho{\mathcal F}) + ({\mathcal F}{\mathcal F} \rho) -\frac{1}{4}(\rho\rho\rho)   \biggr]\,.
\label{2pi}
\eeq
 where $d^4y = \tau' d\tau' d\eta' d^2y_\perp$ and the $\tau$--integral is now defined normally. [For clarity, we omitted indices in the last three terms in the large brackets. They are the same as in the first term. We shall use similar abbreviations below.]
 In the $d^4y$ integral one may prefer to use the retarded Green's function instead of $\rho$
 \beq
 \int^\tau_{\tau_0}d^4y \,\rho(x,y)\cdots = \int^\infty_{\tau_0} d^4y\, iG_R(x,y)\cdots\,.
 \eeq

Similarly,  (\ref{new}) decomposes into the real and imaginary parts
\beq
&& \left[\left( \partial_\tau^2  + \frac{1}{\tau}\partial_\tau  -\frac{1}{\tau^2} \delta_{I\zeta}\right)\delta_{IJ} -(D^2 \delta_{IJ}-D_ID_J-2igF_{IJ})\right]^{ab}{\mathcal F}^{bc}_{JK}(x,y) \nonumber \\
 &&   \qquad + g^2 \left( C_{ad,be} {\mathcal F}^{de}_{IJ}(x,x) + \frac{1}{2} C_{ab,de}{\mathcal F}^{de}_{MM} (x,x)\delta_{IJ} \right) {\mathcal F}^{bc}_{JK}(x,y)
 \nonumber \\
 &&  \qquad \qquad =  -\int^{\tau}_{\tau_0} d^4z\,  \Pi_\rho(x,z)_{IJ}^{ab} {\mathcal F}^{bc}_{JK}(z,y) + \int^{\tau'}_{\tau_0} d^4z\,  \Pi_{\mathcal F}(x,z)_{IJ}^{ab} \rho^{bc}_{JK}(z,y)\,,   \label{real}
\eeq
\beq
&& \left[\left( \partial_\tau^2  + \frac{1}{\tau}\partial_\tau  -\frac{1}{\tau^2} \delta_{I\zeta}\right)\delta_{IJ} -(D^2 \delta_{IJ}-D_ID_J-2igF_{IJ})\right]^{ab}\rho^{bc}_{JK}(x,y) \nonumber \\
 && \qquad   + g^2 \left( C_{ad,be} {\mathcal F}^{de}_{IJ}(x,x) + \frac{1}{2} C_{ab,de}{\mathcal F}^{de}_{MM} (x,x)\delta_{IJ} \right) \rho^{bc}_{JK}(x,y)
 \nonumber \\
 && \qquad \qquad  = - \int^{\tau}_{\tau'} d^4z\, \Pi_\rho(x,z)_{IJ}^{ab} \rho^{bc}_{JK}(z,y)\,,  \label{ima}
\eeq
 where
 \beq
 \Pi_{\mathcal F}(x,y)_{IJ}^{ab} = \frac{1}{2}\, \overrightarrow{V}^x_{alm,ILM}\left({\mathcal F}_{LL'}^{ll'}{\mathcal F}_{MM'}^{mm'}-\frac{1}{4}\rho_{LL'}^{ll'}\rho_{MM'}^{mm'}\right)_{xy}\overleftarrow{V}^y_{bl'm',JL'M'}\,,
 \eeq
and
\beq
\Pi_{\rho}(x,y)_{IJ}^{ab} = \frac{1}{2}\, \overrightarrow{V}^x_{alm,ILM}\left({\mathcal F}_{LL'}^{ll'} \rho_{MM'}^{mm'} +\rho_{LL'}^{ll'} {\mathcal F}_{MM'}^{mm'}\right)_{xy}\overleftarrow{V}^y_{bl'm',JL'M'}\,.  \label{fin}
 \eeq
It is straightforward to include the three--loop contributions to $\Gamma_2$ in the above set of equations. However, the result is rather cumbersome and we do not show them in this paper.

In principle, given the initial condition one can solve these equations numerically and study the degree of thermalization and isotropization by looking at various observables. [Observables should be averaged using the weight function as in (\ref{asin}).] In practice, however, this might be difficult because of the complicated structure of the equations already at two loops. In order to render them  amenable to numerical simulations, in the next section we consider the special case where the classical field $A$ is homogeneous.

\section{The case of the homogeneous background}

\subsection{Evolution equations in the momentum space}

To solve (\ref{2pi}), (\ref{real}) and (\ref{ima}) in full generality is a daunting task as it is computationally too expensive.    The main source of difficulty is the spatial inhomogeneity  of the background field $A_I$. [If we take boost invariance for granted, inhomogeneity here means the dependence on $x_\perp$.] As a matter of fact, in the 2PI literature the problem of solving nonequilibrium evolution equations in the presence of inhomogeneous backgrounds has not been tackled numerically even in scalar field theories.
We thus assume from now on that the background is homogeneous, leaving the inhomogeneous case for future work. From the viewpoint of the application to CGC, admittedly such an assumption is unrealistic because it is in general incompatible with the initial condition (\ref{ini}) $A_i = {\mathcal A}_i^1+{\mathcal A}_i^2$ being dependent on $x_\perp$,  $\partial_i {\mathcal A}_i^{1,2}=\rho^{1,2} \neq 0$. Nevertheless, since the equations are quite nontrivial even after this simplification, one may expect that they still capture the dominant features of thermalization and collective behaviors averaged over  distances larger than $1/Q_s\,$, the typical scale of variation of $A$.

If the background is homogeneous,
we can define the spatial Fourier transform of $Z_{IJ}\equiv \{ {\mathcal F},\rho,\Pi_{\mathcal F},\Pi_\rho\}_{IJ}$ as
 \beq
 Z_{IJ}(\tau,\tau',\eta-\eta',x_\perp-x'_\perp) = \int \frac{d^3p}{(2\pi)^3}
 e^{ip_\eta (\eta-\eta') + ip_\perp (x_\perp-x'_\perp)} Z_{IJ}(\tau,\tau',p)\,,
 \eeq
  where $p_\alpha =(p_\eta,p_\perp)$ and $d^3p=dp_\eta d^2p_\perp$. With this definition ${\mathcal F}(p)$ and $\rho(p)$ are dimensionless while $\Pi_{\mathcal F}(p)$ and $\Pi_{\rho}(p)$ have dimension 4.
Note that we cannot perform the Fourier transform with respect to the rescaled momentum $p_\zeta \leftrightarrow \zeta$.
  Nevertheless, the notation $p_I=(p_\zeta,p_\perp)$ is still convenient. The covariant derivative $D_\zeta$ acting on the left of the Green's function  becomes, in the momentum space,
  \beq
  &&   (\partial_\zeta -igA_\zeta) I(\tau,\tau',\eta-\eta') = \frac{1}{\tau} (\partial_\eta -ig A_\eta) Z(\tau,\tau',\eta-\eta') \nonumber \\
  &&   \qquad \to
  \frac{1}{\tau}(ip_\eta -ig A_\eta) Z(\tau,\tau',p_\eta) = (ip_\zeta -igA_\zeta)Z(\tau,\tau',p_\eta)\,.
  \eeq
   In the following, we use $p_\alpha$ in loop integrals $d^3p$ and in the argument of the Green's function, whereas  we use $p_I$ in the covariant derivative keeping in mind which value of $\tau$ is used to rescale $p_\zeta=p_\eta/\tau$.

 In scalar theories with a homogeneous  background, $Z(\tau,\tau',p)$ is real due to parity. However, in gauge theories this is no longer the case. Since $Z(p)$ implicitly depends on $A$, parity symmetry and the condition that $Z$ is real in the coordinate space
  \beq
  Z(p,A) = Z(-p,-A) = Z^*(-p,A)\,,
  \eeq
  only require  that the real part of $Z(p)$ is even in $p$ while the imaginary part is odd in $p$. Thus, for instance the first term in
 the second line of  (\ref{2pi}) becomes
 \beq
 (\partial^x_I \delta^{ce}+gf_{cde}A_I^d)
 {\mathcal F}_{JJ}^{eb}(x,y)\big\arrowvert_{y=x} = \int \frac{d^3p}{(2\pi)^3} (ip_I \delta^{ce}+gf_{cde}A_I^d(\tau))
 {\mathcal F}_{JJ}^{eb}(\tau,\tau,p)
 \nonumber \\
  =\int \frac{d^3p}{(2\pi)^3} \left(-p_I \,{\mathcal Im}{\mathcal F}_{JJ}^{cb}(\tau,\tau,p)+gf_{cde}A_I^d(\tau)
\, {\mathcal Re} {\mathcal F}_{JJ}^{eb}(\tau,\tau,p)\right)\,.
\eeq
 Similarly, in the cubic terms of (\ref{2pi})
 \beq
&& \frac{ig^2}{2} C_{ab,cd} \int^\tau_{\tau_0} \tau'd\tau' \int  \frac{d^3 p}{(2\pi)^3} \frac{d^3q}{(2\pi)^3}   V_{lmn,LMN}(p,q,\tau') \nonumber \\
&& \times \biggl[ \rho_{IL}^{bl}(\tau,\tau',p){\mathcal F}_{JM}^{cm}(\tau,\tau',q){\mathcal F}_{JN}^{dn}(\tau,\tau',-p-q) +({\mathcal F}\rho{\mathcal F}) + ({\mathcal F}{\mathcal F} \rho) -\frac{1}{4}(\rho\rho\rho)   \biggr]\,, \label{cub}
\eeq
with  the dressed gluon vertex in the momentum space
 \beq
 V_{lmn,LMN}(p,q,\tau)&=& -ig\biggl\{ \delta_{LM}\bigl(i(p_N-q_N )f_{lmn} +gC_{lm,np} A^{p}_N(\tau)\bigr)  \nonumber \\
&&  \qquad + \delta_{LN}\bigl(i(-2p_M-q_M)f_{lmn} +gC_{ln,mp} A^{p}_M(\tau)\bigr)  \nonumber \\
&& \qquad \qquad + \delta_{MN}\bigl(i(p_L+2q_L)f_{lmn} +gC_{mn,lp}A^{p}_L(\tau) \bigr) \biggr\}\,,
\eeq
  the terms in $V$ which are linear in momenta do not vanish even though the background $A$ does not carry spatial momentum.

The equations for the fluctuation (\ref{real}), (\ref{ima}) become manifestly complex--valued
\beq
 &&  \left( \partial_\tau^2  + \frac{1}{\tau}\partial_\tau  -\frac{1}{\tau^2} \delta_{I\zeta}\right) {\mathcal F}^{ac}_{IK}(\tau,\tau',p)  \nonumber \\ &&
  -(ip_L \delta^{ad} +gf_{aed}A_L^e)(ip_L \delta^{db}+gf_{dfb}A_L^f){\mathcal F}^{bc}_{IK}(\tau,\tau',p) \nonumber \\
  &&
  +\left[ (ip_I \delta^{ad}+gf_{aed}A_I^e)(ip_J \delta^{db}+gf_{dfb}A_J^f) + 2g^2 f_{abd}f_{def}
  A_I^e A_J^f  \right] {\mathcal F}_{JK}^{bc}(\tau,\tau',p) \nonumber \\
  &&   \qquad + g^2\int \frac{d^3q}{(2\pi)^3} \left(  C_{ad,be}{\mathcal F}^{de}_{IJ}(\tau,\tau,q) +\frac{1}{2}C_{ab,de} {\mathcal F}^{de}_{MM} (\tau,\tau,q)\delta_{IJ} \right) {\mathcal F}^{bc}_{JK}(\tau,\tau',p)
 \nonumber \\
 &&  =  -\int^{\tau}_{\tau_0} \tau'' d\tau'' \Pi_\rho(\tau,\tau'',p)_{IJ}^{ab} {\mathcal F}^{bc}_{JK}(\tau'',\tau',p) + \int^{\tau'}_{\tau_0} \tau'' d\tau'' \Pi_{\mathcal F}(\tau,\tau'',p)_{IJ}^{ab} \rho^{bc}_{JK}(\tau'',\tau',p)\,, \nonumber
\eeq
 where the self--energy in the momentum space is
\beq
 \Pi_{\mathcal F}(\tau,\tau'',p)_{IJ}^{ab}&=&  \frac{1}{2} \int\frac{d^3q}{(2\pi)^3}V_{amn,IMN}(p,q,\tau) \Bigl(  {\mathcal F}_{MM'}^{mm'}(\tau,\tau'',-q){\mathcal F}_{NN'}^{nn'}(\tau,\tau'',p+q) \nonumber \\
&& \quad  - \frac{1}{4}\rho_{MM'}^{mm'}(\tau,\tau'',-q)\rho_{NN'}^{nn'}(\tau,\tau'',p+q) \Bigr) V_{bm'n',JM'N'}(-p,-q,\tau'')\,, \nonumber
\eeq
 \beq
\Pi_\rho(\tau,\tau'',p)_{IJ}^{ab}&=&  \frac{1}{2} \int\frac{d^3q}{(2\pi)^3}V_{amn,IMN}(p,q,\tau) \Bigl(  {\mathcal F}_{MM'}^{mm'}(\tau,\tau'',-q)\rho_{NN'}^{nn'}(\tau,\tau'',p+q) \nonumber \\
&& \ \   +\rho_{MM'}^{mm'}(\tau,\tau'',-q){\mathcal F}_{NN'}^{nn'}(\tau,\tau'',p+q) \Bigr) V_{bm'n',JM'N'}(-p,-q,\tau'')\,.
\eeq

\subsection{The initial condition}

  Having discussed the evolution equations in the momentum space, we now specify the initial conditions at some initial time $\tau=\tau_0\approx 0$.  The initial conditions for the classical field are already stated  in (\ref{ini}).
  The second equation in (\ref{ini}) implies  $A_\eta \sim {\mathcal O}(\tau^2)$ as $\tau \to 0$, from which the initial condition for $A_\zeta=A_\eta/\tau$ follows
 \beq
 A_\zeta =0, \qquad \partial_\tau A^a_\zeta =-\frac{g}{2}f_{abc}{\mathcal A}_i^{1b}{\mathcal A}_i^{2c}\,.
 \eeq
 On the other hand,  the initial conditions for the fluctuations ${\mathcal F}(\tau_0,\tau_0,p)$  turn out to be nontrivial and have been a subject of recent debate.
 Commonly in the 2PI literature, ${\mathcal F}(\tau_0,\tau_0,p)$ is chosen to be an arbitrary function far from the equilibrium distribution (like the Gaussian or `tsunami' distribution). However, in our problem the initial condition is not arbitrary, but fixed by the underlying QCD Lagrangian.

 Let us first consider the free theory without the classical background.
 The initial distribution of the fluctuation (`Wigner function') derived in \cite{Fukushima:2006ax} is equivalent, in our terminology, to the initial condition for the statistical part of the Green's function.
 The results of \cite{Fukushima:2006ax} translate into
 \beq
&& {\mathcal F}_{ij}^{ab}(\tau_0,\tau_0,p)
 = \delta_{ab} \, \frac{1}{\sqrt{p_\eta^2+\tau_0^2 p_\perp^2}} \left(\delta_{ij}+\frac{\tau_0^2 p_ip_j}{p_\eta^2}\right)
\approx \delta_{ab}\delta_{ij} \frac{1}{|p_\eta|}\,,   \label{0} \\
 && \partial_\tau \partial_{\tau'}{\mathcal F}_{ij}^{ab}(\tau,\tau',p)
 \big\arrowvert_{\tau=\tau'=\tau_0}
=\delta_{ab} \frac{\sqrt{p_\eta^2+\tau_0^2 p_\perp^2}}{\tau_0^2} \left(\delta_{ij}-\frac{\tau_0^2 p_ip_j}{p_\eta^2+\tau_0^2 p_\perp^2}\right) \approx \delta_{ab}\delta_{ij}\frac{|p_\eta|}{\tau^2_0}\,, \label{first}
 \eeq
  where the last expressions are valid when $|p_\eta| \gg \tau_0|p_\perp|$.
   As for the initial condition for ${\mathcal F}_{\zeta\zeta}$, one should note that the longitudinal field $a_\eta = \tau a_\zeta$ and their correlation functions are  constrained by the Gauss's law
  \beq
   \frac{1}{\tau}\partial_\eta \partial_\tau a_\eta = \partial_\eta e^\eta = -D_i e^i = -\tau D_i \partial_\tau a_i\,,  \label{vergauss}
   \eeq
  as  obtained by linearizing (\ref{gauss}) around the classical background $E^i\to E^i+e^i$, $E^\eta\to E^\eta + e^\eta$.
  Using ${\mathcal F}_{\eta\eta} = \tau\tau'{\mathcal F}_{\zeta\zeta} \sim {\mathcal O}(\tau^2\tau'^2)$, one finds, in the absence of the background,
 \beq
  {\mathcal F}_{\zeta\zeta}^{ab}(\tau_0,\tau_0,p)\approx  0\,,   \label{ok}
\eeq
\beq
  \partial_\tau \partial_{\tau'}{\mathcal F}_{\zeta\zeta}^{ab}(\tau,\tau',p)\big\arrowvert_{\tau=\tau'=\tau_0} &\approx&  \frac{1}{4\tau\tau'} \partial_\tau \partial_{\tau'}{\mathcal F}_{\eta\eta}^{ab}(\tau,\tau',p)\big\arrowvert_{\tau=\tau'=\tau_0}  \nonumber \\  &\approx& \frac{\tau\tau'}{4} \frac{p_ip_j}{p_\eta^2} \partial_\tau \partial_{\tau'}{\mathcal F}_{ij}^{ab}(\tau,\tau',p)
 \big\arrowvert_{\tau=\tau'=\tau_0}  \nonumber \\
 &=&\delta_{ab}\frac{p_\perp^2}{4\sqrt{p_\eta^2+\tau_0^2 p_\perp^2}} \approx \delta_{ab} \frac{p_\perp^2}{4|p_\eta|} \,,  \label{free}
  \eeq
  where we assumed $p_\eta \neq 0$, see below.

 The presence of the background $A_i$ will modify the above formulae. Roughly, $p_i \to p_i - gA_i\,$. More precisely, as discussed in \cite{Dusling:2011rz}, one has to diagonalize  the matrix ${\mathcal M}$ in the quadratic form\footnote{If the background $A$ depends on $x_\perp$, this involves the diagonalization of a large matrix with the dimension proportional to the spatial volume.}
 \beq
a^a_i {\mathcal M}_{ij}^{ab} a^b_j \equiv a^{a}_i \bigl[-(D_LD_L)^{ab} \delta_{ij} + (D_i D_j)^{ab} + 2igF_{ij}^{ab} \bigr] a^b_j\,,
\eeq
for each realization of $A$, from which one finds ${\mathcal F} \sim {\mathcal M}^{-1/2}$. Clearly, this procedure will affect only the subleading terms in $\tau_0^2$ so that for our purpose they may be neglected by taking $\tau_0$ arbitrarily  small (except in (\ref{free})). For instance, if we discretize the spatial coordinates $(\eta,x_\perp)$ by putting them in a periodic box of size $(L_\eta,L_\perp)=(2N_\eta a_\eta, 2N_\perp a_\perp)$ where $a_\eta$ and $a_\perp$ are the lattice spacings, the condition $|p_\eta| \gg \tau_0 |p_\perp|, \tau_0 |A_i|$ reduces to
\beq
\frac{2\pi |n_\eta|}{2N_\eta a_\eta}  \gg  \tau_0\,  {\rm max}\left\{\frac{2\pi |n_\perp|}{2N_\perp a_\perp}\,,\   gA_i\right\}\,. \label{wesee}
\eeq
Taking $gA_i \sim Q_s < 1/a_\perp$ and $2N_\eta a_\eta \sim 5$, the typical size of the (approximately) boost invariant region  in heavy--ion collisions, we see that the above condition can be well satisfied even for $n_\eta=1$ if we use a very anisotropic lattice $\tau_0\ll  a_\perp$. Such an anisotropic lattice is commonly used in  the 2PI literature.
   As for the zero mode $n_\eta=p_\eta= 0$, we can neglect it altogether at initial time since $a_i$ is then  independent of $\eta$ and can therefore be absorbed into  the initial background $A_i$ which itself is a random variable in CGC.
We thus conclude that the leading ($\tau_0\to 0$) terms of (\ref{0}) and (\ref{first}) may be used as the initial condition of our equations in the presence of the background, whereas (\ref{free}) is modified by the Gauss's law (\ref{vergauss}) to
\beq
 && \partial_\tau \partial_{\tau'}{\mathcal F}_{\zeta\zeta}^{ab}(\tau,\tau',p)
 \big\arrowvert_{\tau=\tau'=\tau_0} \nonumber \\
&& = \frac{-1}{ 4p_\eta^2}
  (\delta^{ad}p_i  -igf_{acd}A_i^c) (-\delta^{bd}p_j -igf_{bed}A_j^{e})\,   \sqrt{p_\eta^2+\tau_0^2p_\perp^2} \left(\delta_{ij}-\frac{\tau_0^2 p_ip_j}{p_\eta^2+\tau_0^2 p_\perp^2}\right) \nonumber \\
  &&  \approx \frac{1}{4|p_\eta|} (\delta^{ad} p_i -igf_{acd}A_i^c ) (\delta^{bd}p_i +igf_{bed}A_i^{e} )\,.
   \label{second}
 \eeq
For $\tau > \tau_0$, the evolution equations automatically keep terms of order $\tau^2 p^2_i$ and $\tau^2 (gA_i)^2$.

 Finally we derive the initial conditions for the spectral part of the Green's function. The canonical commutation relation (in the momentum space)
  \beq
  [a_i, \tau \partial_\tau a_j] = i\delta_{ij}\,, \qquad
  [a_\eta, \frac{1}{\tau}\partial_\tau a_\eta] =i\,,
  \eeq
  immediately gives
  \beq
  &&\rho_{ij}^{ab}(\tau_0,\tau_0,p)
  =0\,, \qquad \partial_\tau \rho_{ij}^{ab}(\tau,\tau_0)\arrowvert_{\tau=\tau_0} =- \partial_\tau \rho_{ij}^{ab}(\tau_0,\tau)\arrowvert_{\tau=\tau_0} = \frac{1}{\tau_0}\delta_{ij}\delta^{ab}\,, \nonumber \\
  && \rho_{\eta\eta}^{ab}(\tau_0,\tau_0,p)
  = 0\,, \qquad
   \partial_\tau \rho_{\eta\eta}^{ab}(\tau,\tau_0)\arrowvert_{\tau=\tau_0}=- \partial_\tau \rho_{\eta\eta}^{ab}(\tau_0,\tau)\arrowvert_{\tau=\tau_0} =\tau_0 \delta^{ab}\,.  \label{spec}
  \eeq
  Being derived solely from the canonical commutation relation, Eq.~(\ref{spec}) actually holds for all values of $\tau$, and moreover, even in the presence of the background field. The former property will serve as a good check of the numerical simulation.
   Eq.~(\ref{spec}) together with the equation of motion for $\rho_{\eta\eta}$ (as $\tau\to 0$) and the antisymmetric property $\rho_{\eta\eta}(\tau,\tau')=-\rho_{\eta\eta}(\tau',\tau)$ uniquely fix the small--$\tau$ behavior
  \beq
  \rho_{\eta\eta}(\tau,\tau',p) \approx \frac{1}{2}(\tau^2-\tau'^2)\,, \qquad (\tau,\tau'\to 0)\,.
  \eeq
  One thus finds the initial condition for $\rho_{\zeta\zeta}$
  \beq
  \rho_{\zeta\zeta}(\tau,\tau',p) \approx \frac{(\tau^2-\tau'^2)}{2\tau\tau'}\,, \qquad (\tau,\tau'\to 0)\,, \qquad
  \partial_\tau \rho_{\zeta\zeta}^{ab}(\tau,\tau_0)\arrowvert_{\tau=\tau_0} =\frac{1}{\tau_0} \delta^{ab}\,.
  \eeq

\section{Discussions}

\begin{figure}
  \begin{center}
      \includegraphics[width=120mm]{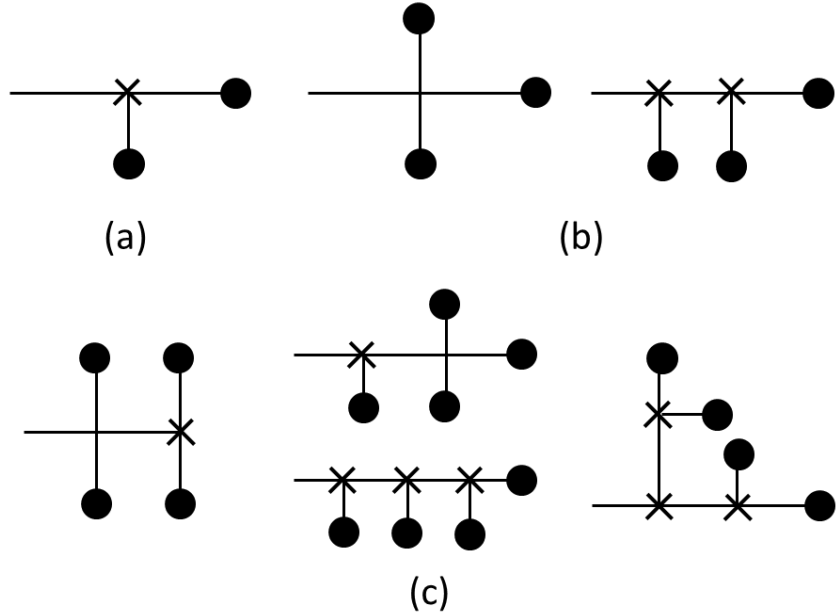}
  \end{center}
  \caption{Tree diagrams obtained by the perturbative expansion (\ref{exp}). Crosses and blobs denote the insertions of $D^{(0)}$ and $A^{(1)}$, respectively. Solid lines connecting vertices are the retarded propagators $G_R^{(0)}\propto \rho^{(0)}$ in the background of $A^{(0)}$.}
  \label{tree}
\end{figure}

\begin{figure}
  \begin{center}
      \includegraphics[width=120mm]{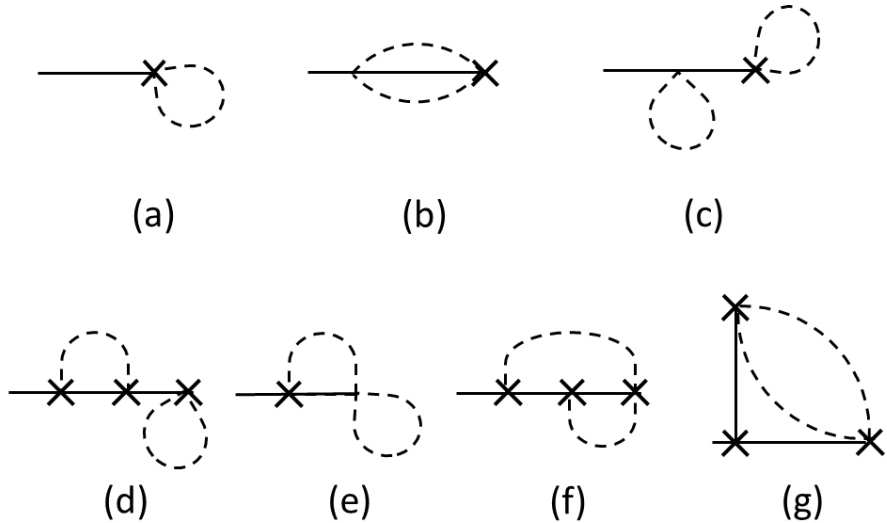}
  \end{center}
  \caption{Loop diagrams obtained from Fig.~\ref{tree} after performing all possible contractions of all the blobs. Dashed lines denote ${\mathcal F}^{(0)}$. }
  \label{loop}
\end{figure}

An efficient and illuminating way to describe the nature of the equations (\ref{2pi}), (\ref{real}) and (\ref{ima}) would be to compare with the previous CGC--based works.
The approach pursued in   \cite{Dusling:2010rm,Dusling:2011rz} is an extension  to QCD of a technique developed for a scalar theory \cite{Son:1996zs}.
One first  solves the Yang--Mills equation (\ref{ym})
perturbatively
\beq
A_I = A_I^{(0)}+A_I^{(1)}+A_I^{(2)}+\cdots \,, \qquad A_I^{(n)} \sim {\mathcal O}(g^{n-1})\,,
\label{exp}
\eeq
where
$A^{(0)}_I$ is itself an exact solution to the Yang--Mills equation and $A^{(1)}_I$ satisfies the linearized equation \beq
\left[\left( \partial_\tau^2  + \frac{1}{\tau}\partial_\tau  -\frac{1}{\tau^2} \delta_{I\zeta}\right)\delta_{IJ} -\left((D^{(0)})^2 \delta_{IJ}-D^{(0)}_ID^{(0)}_J-2igF^{(0)}_{IJ}\right)\right]^{ab}A^{(1)b}_J=0\,.
\eeq
The initial condition is taken to be $A(\tau_0)=A^{(0)}(\tau_0)+A^{(1)}(\tau_0)$.
Since $A^{(0)}+A^{(1)}$ is not an exact solution, one finds the (linearized) equation of motion for $A^{(2)}$
\beq
&& \left[\left( \partial_\tau^2  + \frac{1}{\tau}\partial_\tau  -\frac{1}{\tau^2} \delta_{I\zeta}\right)\delta_{IJ} -\left((D^{(0)})^2 \delta_{IJ}-D^{(0)}_ID^{(0)}_J-2igF^{(0)}_{IJ}\right)\right]^{ab}A^{(2)b}_J \nonumber \\
  && \qquad = gf_{abc} \left[ \left(D_{I}^{be}
 A^{(1)e}_J\right)A^{(1)c}_J +\left(D_{J}^{be}A^{(1)e}_J\right)A^{(1)c}_I
 -2\left(D_{J}^{be}A^{(1)e}_I\right)A^{(1)c}_J \right]\,. \label{sch}
\eeq
Note that the right hand side is identical to the second line of (\ref{2pi}) if one replaces
${\mathcal F}(x,y)\to {\mathcal F}^{(0)}(x,y)\equiv  A^{(1)}(x)A^{(1)}(y)$. Let us write (\ref{sch}) schematically using the retarded Green's function $G_R^{(0)} \propto \rho^{(0)}$ in the background of $A^{(0)}$
\beq
A^{(2)} \sim g \rho^{(0)} D^{(0)}A^{(1)}A^{(1)}\,,
\eeq
 and represent this as in Fig.~\ref{tree}a.  Repeating this procedure, one finds two contributions
 \beq
 A^{(3)} \sim g\rho^{(0)}(gA^{(1)}A^{(1)}A^{(1)}+  D^{(0)}A^{(1)}A^{(2)} ) \,,
 \eeq
 which are depicted by the two diagrams in Fig.~\ref{tree}b, respectively, and
 \beq
 A^{(4)} \sim g \rho^{(0)} (gA^{(1)}A^{(1)}A^{(2)}+D^{(0)}A^{(1)}A^{(3)} + D^{(0)}A^{(2)}A^{(2)})\,,
 \eeq
 corresponding to the  diagrams in Fig.~\ref{tree}c.  Continuing in this way, one generates an infinite number of tree diagrams made up of $D^{(0)}$, $A^{(1)}$ and the retarded propagator. One then makes `contractions' of pairs of $A^{(1)}$ in all possible ways (by using an appropriate Gaussian weight functional of $A^{(1)}(\tau_0)$) and identify each pair with ${\mathcal F}^{(0)}$. As a result of this, one gets  loop diagrams as depicted in Fig.~\ref{loop}. [Note that $A^{(3)}$ vanishes after this operation since it has an odd number of $A^{(1)}$'s.] Each building block of these diagrams has a counterpart in (\ref{2pi}), (\ref{real}) and (\ref{ima}): Diagram 4b corresponds to the terms $\sim \rho{\mathcal F}{\mathcal F}$ in (\ref{2pi}). The tadpole on the retarded propagator in Diagram 4c  represents the modification of the $\rho$--function which is taken into account by the $ {\mathcal F}\rho$ terms in (\ref{ima}). The structure $ \rho^{(0)} {\mathcal F}^{(0)}\rho^{(0)}\rho^{(0)}$ in Diagram 4d corresponds to the $\Pi_\rho\, \rho\sim ({\mathcal F}\rho)\rho$ terms in (\ref{ima}), $\rho^{(0)}{\mathcal F}^{(0)}{\mathcal F}^{(0)}$ in Diagram 4e corresponds to the ${\mathcal F}{\mathcal F}$ terms in (\ref{real}), $\rho^{(0)} \rho^{(0)} {\mathcal F}^{(0)}{\mathcal F}^{(0)}$ of Diagram 4f corresponds to the $\Pi_{\rho}{\mathcal F}\sim (\rho {\mathcal F}){\mathcal F}$ term in (\ref{real}), and $\rho^{(0)}{\mathcal F}^{(0)}{\mathcal F}^{(0)}\rho^{(0)}$ in Diagram 4g corresponds to the $\Pi_{\mathcal F}\rho \sim ({\mathcal F}{\mathcal F})\rho$ term in (\ref{real}).

Higher order diagrams for $A^{(6,8,...)}$ which are solely made up of the above building blocks are automatically resummed by the nonlinear equations (\ref{2pi})--(\ref{ima}). However, diagrams which contain more complicated irreducible structures can only be resummed by evaluating $\Gamma_2$ to three loops (as in (\ref{A})--(\ref{D})) and beyond. These higher order contributions are initially suppressed by powers of the coupling  $A^{(n)}\sim g^{n-1}$, but at late times they may become important since they are multiplied by factors of ${\mathcal F}$ which may grow at low momentum due to some sort of instability (see, e.g., \cite{Kurkela:2011ti} and references therein).

  On the other hand, already at two loops there are terms in (\ref{2pi})--(\ref{ima}) which have no counterpart in the above diagrammatic expansion. These are the $\rho\rho\rho$ term in (\ref{2pi}) and the $\rho\rho$ term in $\Pi_{\mathcal F}$. [There are no loops formed by two retarded propagators.] Neglecting these terms is precisely the content of the classical statistical approximation \cite{Berges:2004yj}. Thus one can view the CGC--based approach \cite{Dusling:2010rm,Dusling:2011rz} as a generalization of the classical statistical approach  \cite{Berges:2007re,Berges:2008mr} to the situation where there is a nonzero background: In the former, one characterizes the large occupation number of gluons by the (inhomogeneous) background $A^{(0)}[\rho]\sim 1/g$ associated with the source  charge $\rho$, while the initial distribution of the fluctuations ${\mathcal F}(\tau_0,\tau_0)$ at $\tau_0 \approx 0$ is uniquely fixed by the QCD Lagrangian expanded to quadratic order.  In the classical statistical approach, one usually sets $A^{(0)}=0$ assuming there is no source (or by gauge invariance), and characterizes the large occupation number by an arbitrary, but physically motivated function ${\mathcal F}(\tau^0,\tau^0)=\langle A^{(1)}(\tau_0)A^{(1)}(\tau_0)\rangle$ according to which the initial configurations $A^{(1)}(\tau_0)$ are generated. This provides an effective description of the evolution after the classical field has decayed substantially, meaning that $\tau_0\gtrsim 1/Q_s$.

 As is well known, the classical statistical approximation is valid only for  low momentum modes where the occupation number is large. In a free massless scalar theory in flat space,
    \beq
    {\mathcal F}(t-t',p) = \frac{\cos (t-t')p}{p}\left(n(p)+\frac{1}{2}\right)\,,
\quad \rho(t-t',p)= \frac{\sin (t-t')p}{p}\,,
\eeq
where $n(p)$ is the Bose--Einstein distribution at the temperature $T$. The condition $|{\mathcal F}|\gg |\rho|$ is valid for momenta much smaller than the temperature $p\ll T$ so that $n(p)\gg 1$, and even in this regime the approximation is valid only in some averaged sense due to the oscillation in $t-t'$. Because of this, the classical approach cannot reproduce the Bose--Einstein distribution, or its  Boltzmann tail. Instead, one finds \cite{Jeon:2004dh,Epelbaum:2011pc}
\beq
n(p)=\frac{T}{p}-\frac{1}{2}\,, \label{condition}
\eeq
as the equilibrium distribution as determined from the vanishing condition of the collision integral $\Pi_\rho {\mathcal F} - \Pi_{\mathcal F} \,\rho$ (c.f., (\ref{real})). The two terms in (\ref{condition}) are precisely the first two terms in the expansion of the Bose--Einstein distribution at $p\ll T$.

 In contrast, in the 2PI formalism  one is  guaranteed to obtain the Bose--Einstein distribution as the equilibrium distribution because by including the $\rho\rho$ terms, or rather, by not neglecting any term one can satisfy the detailed balance within each of the diagrams contributing to $\Pi$. This in particular means that the formalism encompasses a relatively large (ideally, an infinitely large) region in momentum  reaching out to the Boltzmann (exponential) distribution which is an unmistakable feature of heavy--ion collisions. In principle, the present approach has the potential to describe the evolution of the system from right after the collision all the way to the late quantum regime in a single framework.

 In conclusion, we have set up the foundation to incorporate the physics of the CGC  in the 2PI formalism. As already remarked in Section 3, there still remain important formal problems concerning gauge invariance and renormalizability \cite{Arrizabalaga:2002hn,Carrington:2003ut,Reinosa:2007vi,Reinosa:2009tc} which have to be addressed in order to place this approach on firmer ground. And of course, eventually one would like to solve the evolution equations numerically and apply the results to the phenomenology in heavy--ion collisions. At least in the homogeneous case, the causal structure of the equation allows for a straightforward numerical implementation (provided some renormalization scheme) as an initial value problem. Yet, manipulations of large matrices such as ${\mathcal F}_{IJ}^{ab}(\tau,\tau',p_\alpha)$ may  pose serious challenges for limited computer resources. Work in this direction is under way \cite{hatta}.

\vspace{10mm}
\section*{Acknowledgments}
We are grateful to J\"urgen Berges, Fran\c{c}ois Gelis and S\"oren Schlichting for  discussions and critical comments. We also thank Kevin Dusling and Raju Venugopalan for helpful conversation.
 This work is supported by Special Coordination Funds for Promoting Science and Technology
 of the Ministry of Education, the Japanese Government.

\appendix

\section{The 2PI action to three--loops}

In this appendix we consider $\Gamma_2$ to three loops and calculate its functional derivatives.
  There are three diagrams as depicted in Fig.~\ref{f1}c--\ref{f1}e. The last two diagrams give the following contributions to be added to the right hand side of (\ref{origin}).\\
 Diagram \ref{f1}d:
\beq
\frac{\delta \Gamma_2^{(1{\rm d})}}{\delta A_I^a(x)} &=& \frac{-g^2}{2}C_{ab,lm}\int dy dz dw \, \overrightarrow{V}_{pp'r,PP'R}^z G_{II'}^{bb'} (x,y) G_{LP}^{lp}(x,z)G_{LQ}^{mq}(x,w) \nonumber \\
&& \quad \times G_{RR'}^{rr'}(z,w)
G_{P'L'}^{p'l'}(z,y)G_{Q'M'}^{q'm'}(w,y)
\overleftarrow{V}^w_{qq'r',QQ'R'}\overleftarrow{V}^y_{b'l'm',I'L'M'}\,.  \label{1d}
\eeq
Diagram \ref{f1}e:
\beq
\frac{\delta \Gamma_2^{(1{\rm e})}}{\delta A_I^a(x)} &=& \frac{ig^4}{4}C_{ab,ll'} C_{pp',qq'}\int dydz\, V^y_{b'mm',I'MM'}  G_{QM}^{qm}(z,y)
 \nonumber \\ &&
  \times \left(G_{II'}^{bb'}(x,y)G_{LP}^{lp}(x,z)
+2G_{LI'}^{lb'}(x,y)G_{IP}^{bp}(x,z)\right)
\nonumber \\
&& \quad  \times \left(G_{LP}^{l'p'}(x,z)G_{QM'}^{q'm'}(z,y)+2G^{l'q'}_{LQ}(x,z)G_{PM'}^{p'm'}(z,y)
 \right)
 \,.   \label{1e}
\eeq

The two--loop self--energy diagrams coming from Fig.\ref{f1}c--\ref{f1}e are shown in Fig.~\ref{f2}b--2e. They give corrections to (\ref{nonl}) and are evaluated as\\
Diagram 2b:
\beq
(\Pi^{2{\rm b}})_{IJ}^{ab}(x,y) =-\frac{g^4}{2}C_{aa',ll'}C_{bb',mm'}G_{LM}^{lm}(x,y) \biggl(G_{IJ}^{a'b'}G_{LM}^{l'm'}
+2G_{IM}^{a'm'}G_{LJ}^{l'b'}\biggr)_{xy}\,. \label{A}
\eeq
Diagram 2c:
\beq
(\Pi^{2{\rm d}})^{ab}_{IJ}(x,y)&=&\frac{1}{4}\int dzdw \, \overrightarrow{V}^x_{alm,ILM}\overrightarrow{V}^z_{pp'r,PP'R}G_{LP}^{lp}(x,z)G_{MQ}^{mq}(x,w)
\nonumber \\  && \quad \times  G_{RR'}^{rr'}(z,w)G_{P'L'}^{p'l'}(z,y)G_{Q'M'}^{q'm'}(w,y)
\overleftarrow{V}^w_{qq'r',QQ'R'}\overleftarrow{V}^y_{bl'm',JL'M'}\,.
\eeq
Diagram 2d:
\beq
(\Pi^{2{\rm c}})^{ab}_{IJ}(x,y)&=&-\frac{ig^2}{4}C_{pp',qq'}\int dz V^x_{all',ILL'}V^y_{bmm',JMM'}G_{LP}^{lp}(x,z)
\nonumber \\
&& \times \left(G_{L'P}^{l'p'}(x,z)G_{QM'}^{q'm'}(z,y)+2G^{l'q'}_{L'Q}(x,z)G_{PM'}^{p'm'}(z,y)
 \right) G_{QM}^{qm}(z,y)\,.
\eeq
Diagram 2e:
\beq
(\Pi^{2{\rm e}})^{ab}_{IJ}(x,y) &=& -\frac{ig^2}{2}C_{bb',nn'} \int dz\, \overrightarrow{V}^x_{alm,ILM}\overrightarrow{V}^z_{pqr,PQR}G_{LP}^{lp}(x,z) \nonumber \\
&& \quad \times G_{QN}^{qn}(z,y)\left(G_{RN'}^{rn'}(x,z)G_{MJ}^{mb'}(x,y)
+2G_{RJ}^{rb'}(z,y)G_{MN'}^{mn'}(x,y)\right)
\nonumber \\
&& + (aI \leftrightarrow bJ)\,.
\label{D}
\eeq

\end{document}